\documentclass[twocolumn]{article}

\begin{document}

\parskip=2mm
\parindent=0mm

\begin{flushright}
physics/0311037 \\
\end{flushright}

 \vskip 0.5 cm
\begin{center}
{\large\bf Quantum Gravity Phenomenology}
%These notes
%were prepared while working on an invited contribution to
% the November 2003 issue of Physics World, which focused on quantum gravity.
%They intend to give a non-technical introduction (accessible to readers
%from outside quantum gravity) to "Quantum Gravity Phenomenology"
\end{center}
\vskip 1.5 cm
\begin{center}
{\small {\bf Giovanni AMELINO-CAMELIA}}\\
\end{center}
\begin{center}
Dipartimento di Fisica,\\
Universit\`{a} di Roma ``La Sapienza''\\ and INFN Sez.~Roma1\\
P.le Moro 2, 00185 Roma, Italy
\end{center}

\vspace{1cm}
\begin{center}
{\bf ABSTRACT}
\end{center}
I give a brief non-technical review of ``Quantum Gravity Phenomenology" and
in particular I describe some studies which should soon allow to
establish valuable data-based constraints on the short-distance structure
of spacetime.

\newpage
\baselineskip 12pt plus .5pt minus .5pt
\pagenumbering{arabic}

\noindent
The ``quantum-gravity problem" has been studied for more than 70 years,
but we still do not have a single experimental result whose interpretation
requires us to advocate a quantum theory of gravity.
The search of a first case in which data provide some evidence
of a quantum property of spacetime is at this point understandably
characterized by a certain anxiety. Even the status
of this research as a truly scientific endeavor could be questioned
if the present lack of confrontation with experiments persists.
And it does not help that one can still encounter in the literature
(although it is fortunately becoming increasingly rare)
some descriptions of the ``quantum-gravity problem"
that are not of the type that one expects in introducing
a truly scientific problem.
For example, as motivation for reseach in quantum gravity
it is sometimes mentioned
that it is unsatisfactory to have on one side
our present description
of electromagnetic, weak and strong forces, unified within the
Standard Model of particle physics, which is a
quantum field theory,
and on the other
side gravity described by General Relativity, which is governed
by the very different rules of classical mechanics.

This type of ``human discomfort", this type of urgency to
correct the lack of elegance of our present worldview,
does not of course define a scientific problem.
But there is a well-defined scientific
problem which can be naturally called ``quantum-gravity problem".
This is the problem of obtaining quantitative predictions for
processes in which both gravity effects
and Standard-Model effects cannot be neglected.
The (quantum) Standard Model of particle physics has been hugely successful
in describing the microscopic phenomena involving fundamental particles,
where gravity can be ignored,
and (classical) General Relativity has been equally successful
in describing the motions of macroscopic bodies, whose quantum
and Standard-Model
properties can be safely neglected.
We have no available data on situations in which neither can be
neglected, but we already know that we would not be able to describe
those data with our current theories, because some logical and mathematical
inconsistecies are encountered even before getting to the point
of obtaining a numerical prediction for these processes.
Like two pieces of different jigsaw puzzles
General Relativity and the Standard
Model cannot be merged without modification.

What would happen if we managed to collide two photons
each with energy of, say, $10^{50}eV$ ?
This is a typical question to which we are unable to provide
even a tentative answer.  According
to our present theories there should not be anything peculiar
in the  {\underline{setup}} of such a collision (just a higher-energy version
of the collisions we already
set up at CERN and other laboratories), but then those same theories
do not allow us to obtain a consistent prediction for the {\underline{outcome}}
of the collision. One might argue that $10^{50}eV$ photons should be the
least of our concerns, since we are never going to be able to produce
and/or observe them, but in cosmology there are some key issues
which require the understanding of ultra-high-energy processes.
Moreover, the fact that our theories fail to generate consistent
predictions in some hard-to-produce contexts
makes us concerned in general
about the robustness of these theories.
Since we know that new elements would have to be introduced in our
theories for the description of ultra-high-energy processes,
it is natural then to wonder whether those new elements can affect
also some contexts in which our present theories do appear
to make sense, contexts that perhaps are more reacheable
than, say, the one of $10^{50}eV$ photons.

The fact that we know so little about the quantum-gravity problem
has a simple explanation. One of the few (perhaps the only)
relatively robust hints that we have about quantum gravity
is that our conventional description of spacetime should start
to break down at the Planck length $L_p \sim 10^{-35} m$
(which can be also described as the inverse of
the Planck energy scale $E_p \sim 10^{28}eV$).
At this scale, gravitational interactions between particles
are no longer negligible. And, based on our experiences in other
similar situations, we expect that this scale should
also dictate how big the effects in the new quantum-gravity theory
will be: for example in processes involving two particles both with
wavelength $\lambda$ the magnitude of the new
effects should be set by some power of the ratio between the Planck length
and the characteristic wavelength of the process $L_p/\lambda$.
Since in all cases accessible to us experimentally $L_p/\lambda$
is extremely small, it is not surprising that the
quantum-gravity effects are nearly inevitably negligible.

While we lack any experimental insight on the quantum-gravity
problem, it is undeniable that quantum-gravity research
has produced some results that appear to be relevant
for the solution of the problem. If nothing else, some theory results
over the last few decades have changed significantly our intuition.
For example, results obtained in string theory~\cite{sussPW}
provide encouragement
for the idea that quantum gravity
could admit a perturbative treatment,
at least in certain contexts.
Before these string-theory results it appeared that quantum gravity
should in all cases be treated using (to-be-determined)
nonperturbative techniques,
with obvious associated difficulties.
The Loop Quantum Gravity~\cite{leePWandBOOK,carloPW} research programme
has looked at the problem from a complementary perspective,
and its development provides encouragement for the idea that a
truly background-spacetime-independent
quantum theory can be constructed. Before these studies
it appeared that there would be a more profound conflict between the
background-spacetime independence of general relativity and
the fact that quantum field theory assumes from the start a background
spacetime.
Another example of a change of intuition comes from
research on certain types of noncommutative
spacetimes (see later in these notes and
Refs.~\cite{doplNCSToldANDnew,majrue,lukieAP,gacmaj,wessNCST,douglasNCST}),
which were the first examples of quantum spacetimes in which
it appeared plausible that the Planck scale might affect
some of the familiar spacetime symmetries, most notably Lorentz symmetry.

Unfortunately, Loop Quantum Gravity can only be treated (so far)
as a fundamentally nonperturbative theory, without access to the tools
of perturbative analysis which are so valuable when we attempt to obtain
quantitative predictions. And on the other hand (in spite of recent
advances~\cite{sussPW}) we do not have
a genuinely background-independent formulation of string theory
(which, I expect, should introduce an {\it a priori}
limitation to the class of
conceivable measurement procedures
for which string theory can provide consistent predictions).
The development of approaches to the quantum-gravity problem based
on noncommutative geometry is at an even earlier
stage of development, since we do not even have, at present, a
compelling candidate for ``noncommutativegeometrodynamics" (a
generalization of  General Relativity that could apply to
noncommutative geometries).
Some of these (and other) unsolved issues
are present in all approaches to the quantum-gravity problem,
and of course the fact that these approaches
have had no experimental success so far
keeps them in the limbo of purely theoretical speculations.
But finally we have some quantum-gravity
approaches which are developed to a point where it
appears plausible that they might have managed
to capture at least some features of the correct theory.
The history of the development of physics tells us that in many
cases theories which ended up not giving a full solution
of the problem of interest nevertheless managed to provide
the right intuition for certain aspects of the full solution.

In attempting to profit from these hints coming from theory
we are of course faced with the mentioned problem of the smallness
of the expected effects, due to the smallness of the Planck length.
For decades it had been assumed that such small effects
could never be seen,
but recently some strategies for overcoming the challenges posed by the
smallness of the effects have been developed.
The type of strategy which is used in
this ``Quantum Gravity Phenomenology"~\cite{polonpapANDqgpIJrev}
can be described in analogy with certain experimental studies
of proton stability.
The prediction of proton decay in grandunified theories of
particle physics
is really a small effect, suppressed by the fourth
power of the ratio between the mass of the proton
and the grandunification scale (an energy scale which should be
only some three orders of magnitude smaller than the Planck energy scale).
In spite of this horrifying suppression,
with a simple idea we have managed to acquire remarkable sensitivity
to the new effect: we keep under observation
a large number of protons, so that, although the probability of
decay of any given proton is very small, the probability that
one of the many monitored protons would decay is observably large.

Analogously in Quantum Gravity Phenomenology
we should focus
on experiments which probe spacetime structure
and host an ordinary-physics dimensionless quantity
large enough that (like the number of monitored protons
in proton-stability studies)
it could amplify the extremely small effects we hope to discover.
And, using this strategy, over the last few years several research
lines within Quantum Gravity Phenomenology have been developed.

Perhaps the most exciting of these research lines are the ones
connected with the idea of a ``quantum spacetime".
In unifying general relativity and quantum mechanics
it is of course natural to contemplate the
idea that space-time itself is quantized.
In ordinary quantum mechanics
the space that contains the three components
of a particle's angular momentum is a ``quantum space":
one cannot make ``sharp" (error-free)
measurements of more than one of the components
of angular momentum (``angular-momentum non-commutativity"),
and a sharp measurement of a single
component can take only certain discrete
values (``angular-momentum discretization").
Similarly one may expect that quantum gravity should lead
to spacetime noncommutativity and/or spacetime discretization.
Space-time non-commutativity would,
in its simplest form, prevent one from making error-free
measurements of more than one of the coordinates of a
point in spacetime. And if spacetime were discrete
certain sharp spacetime measurements (of, say, volume
or one of a point's coordinates) could then take only
certain discrete values.

String Theory does not necessarily lead to spacetime quantization,
at least in the sense that its background spacetime
is commonly taken as a classical geometry.
I fear that the assumption of a classical background spacetime
may point to an incompleteness of String Theory.
In fact, the formalism eventually
leads to the emergence of a fundamental limitation on
the localization of a spacetime event
(see, {\it e.g.}, Refs.~\cite{wittenPT,venezianoetal}),
%(1996 E.~Witten, Phys.~Today 49, 24)
and this might be in conflict
with the assumption of a physically-meaningful classical
background spacetime (whose points acquire operative meaning
only if they can be sharply localized).
In any case (even assuming
that a classical spacetime background should be
acceptable for String Theory) it has been found that under appropriate
conditions (a vacuum expectation value for certain tensor fields)
it is appropriate to set up an effective description
in terms of a noncommutative spacetime~\cite{douglasNCST}.
% (2001 N.R.~Douglas and N.A.~Nekrasov, Rev.~Mod.~Phys.~73, 977).

In Loop Quantum Gravity spacetime is inherently discretized~\cite{lqpdiscr}
%(1995 C.~Rovelli and L.~Smolin, Nucl.~Phys.~B442, 593),
and this occurs in a rather compelling way: it is not that one introduces
by hand an {\it a priori} discrete background spacetime; it is rather
a case in which a background-independent analysis ultimately
leads, by a sort of self-consistency, to the emergence of
spacetime discretization.
Moreover, some recent results~\cite{kodadsr,jurekodadsr}
%(2003 G.~Amelino-Camelia, L.~Smolin and A.~Starodubtsev,
%arXiv.org/hep-th/0306134).
could open the way for a role for noncommutative spacetimes
also in the description of some aspects of
Loop Quantum Gravity.

A description of space-time that is ``fundamentally quantum" (quantized,
in the sense of noncommutativity and/or discretization,
even at the level of the background geometry) would require a profound
renewal of fundamental physics. In particular, it should naturally have
striking implications for the  propagation of particles. As usual
in quantum gravity, we expect that these effects would
be very small, characterized by the Planck scale.
A useful analogy is the one of the surface
of a wooden table.  We usually perceive the surface
of the table as perfectly flat, but this flatness
is the result
of some sort of averaging over the short-distance irregularities
of the table's surface. If we take a ball with diameter much larger
than the short-distance scale of roughness of the table surface and roll
it over the surface, we find no evidence of the roughness, but if
we use a ball whose diameter is not  much larger than the scale
of roughness we will see disturbances in the path of the ball
due to the roughness.
Similarly, a quantization of spacetime at the Planck scale
can affect the propagation of particles, in a
way that depends on the ``size" (wavelength) of the particles.
In several recently-studied scenarios for spacetime quantization
(most notably in Loop Quantum Gravity~\cite{leePWandBOOK,gampul}
and in noncommutative
spacetimes~\cite{majrue,lukieAP,gacmaj,wessNCST,douglasNCST}),
it has been shown that the ``dispersion relation" between energy, $E$,
and momentum, $p$, acquires energy/wavelength-dependent
corrections: instead of the familiar
classical-spacetime formula, $E^2 = p^2 + m^2$,
one often finds\footnote{While familiar contexts in which
one encounters
deformed dispersion relations (such as the case of the study
of the propagation of light in certain crystals)
are such that there is a preferred frame, in quantum gravity it
is conceivable that such dispersion relations would emerge
as observer-independent features of spacetime structure~\cite{dsrall}.
I attempted to discuss this point nontechnically in
the recent paper in Ref.~\cite{dsrNATURE}.}
a formula of the type  $E^2 +  \eta E^{2+n}/E_p^n = p^2 + m^2$
(where $\eta$ is of order $1$ and $n$ is an integer).

One context in which spacetime quantization can have
observably large implications is the one of ultra-high-energy
cosmic rays.
According to the present understanding,
cosmic rays are emitted by distant active galaxies and
produce showers of elementary particles when they
pass through the Earth's atmosphere.
Before reaching our Earth detectors
cosmic rays travel gigantic (cosmological) distances and
could interact with photons in the cosmic microwave
background. These interactions,
which can cause energy loss through photopion production,
are kinematically forbidden
at low energies, but they are very efficient when the cosmic-ray energy
is above the photopion-production energy threshold $E_{th}$.
This leads to the expectation that by the time they reach the Earth
a cosmic ray should have energy which does not exceed
significantly $E_{th}$.
However, the estimate of $E_{th}$ depends very sensitively on the
structure of the dispersion relation. Even a minute Planck-scale
modification of the dispersion relation can induce a significant
shift of the cosmic-ray threshold.
Adopting the familiar $E^2 = p^2 + m^2$ one obtains an estimate of
the cosmic-ray threshold that appears to be too low in comparison with
the observations reported by the AGASA cosmic-ray observatory in Japan.
Planck-scale modified dispersion relations can instead
naturally fit~\cite{kifu,grillo,gactprd}
%(1999 T.~Kifune, Astrophys.~J.~Lett.~518, L21).
the AGASA data.

There are other plausible explanations for the AGASA ``cosmic-ray puzzle",
and even on the experimental side the situation must be considered
as preliminary while waiting for more powerful observatories,
such as Auger, in Argentina.
Still, the possibility that in these cosmic-ray studies
we might be witnessing the first manifestation of a quantum
property of spacetime is of course very exciting;
moreover, whether or not they end up being successful
in describing cosmic-ray observations,
these analyses provide an explicit example of a minute Planck-scale
effect that can leave observable traces in actual data.

Besides the cosmic-ray threshold, another key
implication of Planck-scale-modified dispersion relations
is the prediction that the
time needed by a massless particle to propagate over a
given distance should depend on the wavelength/energy of
the particle.
It has been established~\cite{grbgac,billerEschaef,gampul}
%(G.~Amelino-Camelia {\it et al}, Nature 393, 763)
that observations on gamma-ray bursts can be used to
look for this effect.
Gamma-ray bursts provide us clusters of photons emitted by the
source roughly simultaneously, with millisecond
uncertainty\footnote{Here I am actually thinking of a
microburst, within the whole gamma-ray-burst event~\cite{grbgac}.}.
If there was even a tiny (Planck-scale suppressed, as usual)
dependence of the speed of photons on their wavelength
the fact that these photons travel over huge cosmological distances
would lead to a significant correlation between time of arrival
on Earth and wavelength (a correlation that, for photons with energies
higher than $10 MeV$, could lead to an effect which is significant
in comparison with the millisecond
simultaneity available at the source).
The next generation of gamma-ray telescopes, most notably
the GLAST space telescope,
should allow a noteworthy investigation~\cite{glast}
of this scenario.

These studies of Planck-scale deformations of the dispersion
relation (studies based on observations of cosmic rays and gamma-ray bursts)
constitute the best developed research line
in quantum-gravity phenomenology, but there are several other
topics which are under intense investigation. One noteworthy example
are studies of the popular quantum-gravity
intuition  that at a fundamental level the concept of distance
should be  ``fuzzy", it
should be affected by unavoidable quantum fluctuations.
Spacetime noncommutativity was introduced also as a possible way to describe
mathematically this intuition, but
it remains extremely hard to deduce from a given proposal of spacetime
noncommutativity the associated picture of spacetime fuzziness.
In Loop Quantum Gravity it is natural to expect such quantum fluctuations
(whereas in String Theory this point might be affected
by the {\it a priori} assumption of a classical background spacetime),
but again there is still no success in characterizing the effect
in a way that is useful for experiments.

This impasse on the theoretical front of course does not prevent
us from pursuing the issue experimentally, although
the phenomenology of course pays a price for the lack of guidance
from fundamental theories.
Phenomenology of ``distance fuzziness" has primarily focused on its
possible implications for interferometry.
Conventional interferometry relies on the assumption that
there is a reference sharp (classical) concept of length of
the arms of the interferometer.
Preliminary studies in which the classical length is replaced by
a ``fuzzy" length have considered both
atom interferometers~\cite{peri}
%(1998 I.C.~Perival and W.T.~Strunz,Proc.~R.~Soc.~A453, 431)
and laser-light interferometers~\cite{gacgwiALL,ngGWi}.
%(1999 G.~Amelino-Camelia, Nature 398, 216).
These studies showed that it is not unconceivable
that small Planck-scale fuzzy-distance
effects would be seen in the foreseeable future, but since
we are limited to purely phenomenological considerations (because
of the mentioned lack of guidance from more developed quantum-gravity
theories) it is difficult to have an intuition for
the robustness of the estimates.

Related analyses based on spacetime fuzziness, and the associated
subject of quantum-gravity-induced decoherence,
can also be performed in the context of neutral-kaon experiments~\cite{cptALL}.
%(1999 A.~Apostolakis {\it et al}, Phys.~Lett.~B452, 425).
Instead of the high precision of interferometry, these studies
exploit an experimental technique in which a crucial role is played by
the remarkable smallness of the mass difference between the long-lived
and the short-lived neutral kaons. Also in this case
purely-phenomenological sensitivity analyses lead to tentatively
encouraging estimates, but so far there is no evidence
of the effect in data~\cite{nickcpt}.

The typical exercise of the ``quantum-gravity-phenomenology community"
is a situation in which the smallness of the Planck scale
is the key challenge. But I should stress here that recently there
has been some work on the possibility that quantum gravity might
show up at lower energy scales.
There is no good reason for this to happen, but it is
a possibility, and of course it should be explored.
In particular, over these past few years there has  been strong interest
in experiments looking for a key feature of String Theory: the fact that
its mathematical consistency requires the existence of some
extra spacetime dimensions,
in addition to the four we experience directly.
It turns out to be rather natural to assume that these extra
dimensions be of finite Planck-length size.
For this most studied scenario for String Theory, in which the extra
dimensions are, as expected for typical quantum-gravity effects,
characterized by the Planck length, all attempts
to find opportunities for experimental tests have failed.
However, over these past few years there has been interest~\cite{led}
in the possibility
that some of the extra dimensions might have a size which is much larger
than the Planck length.
The most studied scenario assumes two large extra dimensions
and has been shown to lead to important effects already at the
level of {\underline{classical}} gravity, effects which
would be significant
in experimental studies of Newton's $1/r^2$ law when the distance $r$
between two test masses is at or below the $10^{-4} m$ scale.
Experiments looking for these effects adopt various strategies.
For example, Newtonian gravity predicts that there is no net force
between a cylinder and a cylindrical shell that surrounds it,
but any deviation from the inverse-square law would give rise
to a net force.
Experiments looking for such a force have improved significantly
over the last few years, but we still have no evidence
of departures from Newton's inverse-square law even for
distances down to $10^{-4} m$~\cite{newtontest}.
It is amusing that these {\underline{classical}}-gravity tests can
provide an important hint on the {\underline{quantum}}-gravity problem,
since the discovery of extra dimensions would of course be of
encouragement for String Theory, which is the only known theory
that requires extra dimensions for its consistency.

In closing I want to stress that
finally an experimental programme for quantum gravity is
starting to be developed, but it remains extremely difficult.
The fact that String Theory and Loop Quantum Gravity
have matured to the point of providing us some guidance concerning
the nature of the effects that might characterize quantum gravity,
and the recent advances in noncommutative geometry (both as a fundamental
quantum picture of spacetime and as an effective-theory description
of some aspects of String Theory and Loop Quantum Gravity)
have been very valuable for Quantum Gravity Phenomenology.
It is difficult to even attempt guessing where we might be in 15 or 20
years.
By then we will know, through studies of cosmic rays
and gamma-ray bursts, whether or not there is
a Planck-scale modified dispersion relation.
If that response is negative perhaps one of the other reasearch lines
I here mentioned will stumble upon the much needed first experimental
evidence of a quantum property of spacetime.
As usual in phenomenology, no promises can be made concerning the time
needed for a first discovery.
But one goal that Quantum Gravity Phenomenology
will surely achieve in the short term is the one of forcing the
quantum-gravity debate to focus on the potentially observable aspects
of any given quantum-gravity theory.
For too long this research had been in a limbo at the interface
between physics, mathematics and phylosophy.
The excuse for avoiding the difficulties of
a more genuinely scientific attitude toward
the problem was provided by the smallness of the Planck length.
Now that we are planning the first few experiments with
genuine Planck-length sensitivity,
we can no longer hang on to our old excuse.

\section*{Further Reading}
These notes
were prepared while working on an invited contribution to
vol.16 no.11 (November 2003 issue) of Physics World,
which focused on quantum gravity.
While I here focused on ``Quantum Gravity Phenomenology",
other areas of quantum-gravity research were discussed
in Physics World vol.16 no.11 (2003)
29-34 (by L.~Susskind) and Physics World vol.16 no.11 (2003)
37-41 (by C.~Rovelli).
Ref.~\cite{furtherQG} contains a few recent descriptions
of the status of quantum-gravity research.
Ref.~\cite{furtherQGP} contains
a few papers on some strategies for
searching for Planck-scale effects which I did not discuss in
these short notes.
Ref.~\cite{dsrcosmo} contains
some references on the possible role
of Planck-scale-modified dispersion relations in cosmology.

\baselineskip 12pt plus .5pt minus .5pt

\end{document}